# Spin-orbit torque: Moving towards two-dimensional van der Waals heterostructures

R. C. Sahoo[1,2], Dinh Loc Duong[1,3], Jungbum Yoon[1], Pham Nam Hai[4,5,6,*] and Young Hee Lee[1,3,7,*]

[1]Center for Integrated Nanostructure Physics (CINAP), Institute for Basic Science (IBS), Suwon 16419, South Korea
[2]Department of Chemical Science and Engineering, Tokyo Institute of Technology, 2-12-1 Ookayama, Meguro, Tokyo 152-8552, Japan
[3]Department of Energy Science, Sungkyunkwan University, Suwon 16419, South Korea
[4]Department of Electrical and Electronic Engineering, Tokyo Institute of Technology, 2-12-1 Ookayama, Meguro, Tokyo 152-0033, Japan
[5]Center for Spintronics Research Network (CSRN), The University of Tokyo, 7-3-1 Hongo, Bunkyo, Tokyo 113-8656, Japan
[6]CREST, Japan Science and Technology Agency, 4-1-8 Honcho, Kawaguchi, Saitama 332-0012, Japan
[7]Department of Physics, Sungkyunkwan University, Suwon 16419, South Korea

[*)] **Electronic addresses:** *pham.n.ab@m.titech.ac.jp and leeyoung@skku.edu*

**Abstract:**

**The manipulation of magnetic properties using either electrical currents or gate bias is the key of future high-impact nanospintronics applications such as spin-valve read heads, non-volatile logic, and random-access memories. The current technology for magnetic switching with spin-transfer torque requires high current densities, whereas gate-tunable magnetic materials such as ferromagnetic semiconductors and multiferroic materials are still far from practical applications. Recently, magnetic switching induced by pure spin currents using the spin Hall and Rashba effects in heavy metals, called spin-orbit torque (SOT), has emerged as a candidate for designing next-generation magnetic memory with low current densities. The recent discovery of topological materials and two-dimensional (2D) van der Waals (vdW) materials provides opportunities to explore versatile 3D-2D and 2D-2D heterostructures with interesting characteristics. In this review, we introduce the emerging approaches to realizing SOT nanodevices including techniques to evaluate the SOT efficiency as well as the opportunities and challenges of using 2D topological materials and vdW materials in such applications.**

**Contents**

## 1. Introduction

The discovery of spin ordering modulation by means of carrier injection has paved the way for spin-based electronics since 1857 [1]. Spin-based electronics received a second-boost with the revolutionary findings of giant magnetoresistance (GMR) and tunneling magnetoresistance (TMR), where the magnetic states of nanoscale elements can be accurately determined from their electrical resistance in the presence of an external magnetic field [2–5]. The replacement of existing inductive read heads by GMR and TMR hard-drive read heads in data storage has resulted in breakthroughs in storage capacity and design of random-access memory [2,5–8].

The electrical resistance in GMR and TMR-based devices can be efficiently tuned by modulating the magnetization directions of the materials. However, an external magnetic field ($H_{ext}$), created by currents flowing through a nearby conductor, is required to control the material magnetization in standard GMR and TMR memories. This limits the ability to scale down to nanosized devices [5,9,10]. In this scenario, one of the easiest and most effective ways to actively manipulate magnetization of magnetic layers is to use a spin-polarized current through spin angular momentum transfer from the charge current, called spin-torque. The spin angular momentum of the electron current exerts an effective field and torque on the local magnetization of nanoscale-thick magnetic layers. The performance of GMR and TMR-based devices has recently been improved by applying the spin-torque technology. The spin-torque-induced magnetization dynamics are investigated primarily using microlithography in various multilayers, especially nanopillar structures such as Co/Cu [11], Co/Cu/NiFeCo/Cu [12], NiFeCo/Cu/CoFe/Cu [13], and Co/Cu/Co [14]. In these spin-transfer torque (STT) devices, which consist of two ferromagnetic (FM) layers sandwiched by a non-magnetic (NM) thin layer, the spins of one FM layer experience a spin-torque due to angular momentum transfer from the other FM layer when spin-polarized electrons scatter at the interface [15,16]. However, this configuration requires a high current density because of the low spin-transfer efficiency, which induces significant Joule-heating energy loss [17–19]. An emerging approach that uses the lateral current in FM/NM bilayer structures without an additional magnetic layer to create spin-polarized currents called spin-orbit torque (SOT), can be a promising solution for STT devices. It provides further opportunities for future nanospintronics through the enhanced efficiency of torque transfer between the FM and NM or torque layers [20].

Meanwhile, the efficiency of SOT devices can be further improved by choosing appropriate materials of ferromagnetic or non-magnetic layers. Such devices using three-dimensional (3D) materials have a long history but have a few drawbacks such as low efficiency, single dominating component of SOT on magnetization switching, and Joule heating or thermal fluctuation [21–24]. These limitations should be overcome for a desired low power- and energy-efficient SOT devices for realistic applications. An alternative is to use topological insulators (TIs) and van der Waals (vdWs) layered materials. Recently, TIs have been proposed as a promising candidate for efficient torque layer in SOT devices due to the unique behavior of charge carriers [25,26]. They have spin-momentum locking feature in which the spin of the protected topological surface states lies in in-plane and is locked at right angles to the momentum of carriers (Fig. 1a). This feature results in the preferential generation of pure spin current in the direction perpendicular to the film plane [27–29]. Consequently, an in-plane charge current flowing on the surface of TI can effectively generate a spin current perpendicular to the topological surface (Fig. 1b) [30]. This spin current can exert a spin-torque on the adjacent FM layer for magnetization switching. The improvement of SOT efficiency of TI-based devices strongly depends on the carrier density of topological surface states as well as the position of Fermi level, which can be tuned by the impurity doping in $Bi_2Te_3$ family (Fig. 1c) [31,32]. Furthermore, SOT efficiency can be enhanced by modulating the layer thickness as well as applying external gate bias in TIs (Fig. 1d) [33,34].

Another promising classes of SOT materials are two-dimensional (2D) vdWs materials, which have been demonstrated recently. The 2D materials with sufficiently low crystal symmetry [35–37], crystal strain and strong Rashba or spin Hall-type SOC [38,39] are efficient to generate strong SOT, which is much energy resourceful compared to 3D materials. One of the limits of SOT applications (i.e., Joule heating or thermal fluctuation) can also be reduced by using 2D vdWs in monolayer limit. Additional advantages of 2D vdWs materials are their strong SOC in atomically-thin layer [40], perpendicular magnetic anisotropy [41,42], tunable conductivity of topologically protected surface states and easy device fabrication. The intrinsic SOT in 2D vdWs (Fig. 1e), which strongly depend on the SOC-controlled electron spin, can be tuned by modulating crystal symmetry [38], external electric field [43] and proximity effect with transition metal dichalcogenides (TMDs) [44]. The SOC in graphene, for example, can be enhanced by the proximity effect of $WS_2$ [44]. Furthermore, the magnetism of the 2D FM layer can be modulated by different approaches such as impurity doping (Fig. 1f), electrostatic gating (Fig. 1g), and layer thickness [45–47], rather than the traditional magnetic field. For example, magnetism can be introduced or tuned deliberately by magnetic dopants in monolayer semiconductors (e.g. V-doped $WSe_2$, Fe-doped $SnS_2$) [45,48], or by electro-gating using a liquid electrolyte in monolayer $Fe_3GeTe_2$ [47]. Layer-dependent magnetic properties are also observed in $CrI_3$ [49,50], further implying the dimensionality effect in the 2D vdW materials. In the 3D, the distance between dopants is proportional to $1/n^{1/3}$, whereas it is proportional to $1/n^{1/2}$ in the 2D (n stands for percent concentration of dopants) [51]. The difference in the scaling rule between 2D and 3D gives rise to different interaction strength between dopants in 2D and 3D forms at a given doping concentration. Furthermore, the *pd-d* hybridization in 2D TMD DMSs may be more advantageous in applications than the weak *sp-d* hybridization in III-V semiconductors (e.g. Mn-doped GaAs) [52]. The stronger *pd-d* hybridization presents opportunities to overcome the $T_c$ limit in 3D semiconductors. Meanwhile, proximity exchange interactions between 2D-FMs and 2D-SOC play a key role to improve SOT strength in 2D-based SOT devices (Fig. 1h) [53].

Appearing as an emerging field, SOT phenomenon and devices were reviewed extensively from basic principles, experimental methods, and materials [37,54–57]. However, the quick explorations and developments in this research field, especially in topological insulators and 2D materials [31–34,37,57], require a comprehensive updated review. Here, our purpose of this review article is to bring the basic advantages of 2D materials such as topological insulators and van der Waals materials, operation principle of spin-orbit torque in general, the physical mechanism behind it. The three well-known methods to probe SOT efficiency are introduced. We summarize the state-of-art SOT efficiency of different materials for spin-orbit torque devices and explain the reasons why topological insulator and 2D van der Waals materials are promising candidates for improving the spin-orbit torque devices. In the end, we list some challenges and perspectives of using topological insulators and 2D materials for this research field.

## 2. Spin-torque mechanism

Two important phenomena in generating the spin-torque at the interface of the materials are STTs and SOTs. Both phenomena are explained by the similar mechanism, although the configurations of the device structures are different. The key mechanism is the conservation of angular momentum during the scattering process of spin-polarized electrons in magnetic materials. Consequently, the spin-torque is generated and transferred to the adjacent materials, which allows for efficient switching of the magnetic moment. The efficient spin-torque mechanism using the least power consumption is the key to modern nanosprintronics. *We* will discuss each of these spin-torque mechanisms in *detail* below.

### *2.1 Spin-transfer torque*

A typical structure of the STT device consists FM/NM/FM trilayer heterostructure (Fig. 2a) [19,58,59]. Here, the two FM layers have non-collinear magnetizations. The magnetization of the first FM layer is fixed ($\boldsymbol{M}_{\text{fixed}}$), whereas the second FM layer has a free magnetization ($\boldsymbol{M}_{\text{free}}$), which can be easily switched. A spin-unpolarized charge current, $j_c$ (red arrow), flows from the free to the fixed FM layer (right to left) in the FM/NM/FM STT structure, which is equivalent to an electron flow from the fixed to the free FM layer (left to right). After passing through the first ($\boldsymbol{M}_{\text{fixed}}$) FM layer ($FM_{\text{fixed}}$), the unpolarized electrons become polarized along the magnetization direction of $FM_{\text{fixed}}$ layer. The spin-polarized electrons are injected into the second (switchable) FM layer ($FM_{\text{free}}$) through the NM layer (tunnel barrier). This spin-polarized current contains an angular momentum $j_s = (\hbar/2e)p_c j_c$ (where $p_c=(n_{up}-n_{down})/(n_{up}+n_{down})$ is the net polarization of conduction electrons) and undergoes spin-dependent scattering processes [60]. Because the total angular momentum (electron+lattice) is conserved in these scattering processes, a transverse component of the spin angular momentum is ejected into the free FM layer ($FM_{free}$), exerting a torque on $\boldsymbol{M}_{free}$ to conserve angular momentum [15,60]. This spin-torque forcefully changes the orientation of $\boldsymbol{M}_{free}$ when the charge current reaches a critical limit. Therefore, a stable parallel spin configuration between the two FM layers is achieved. The switching efficiency depends on the magnitude ratio of $j_s$ to $M_{free}$ as well as the thickness ($t_{free}$) of the free layer, i.e. $j_s/(M_{free}\times t_{free})$ or $(j_c\times p_c)/(M_{free}\times t_{free})$ [60]. Obviously, a high $p_c$ and a thin $t_{free}$ will give rise to a high efficiency of angular momentum transfer.

When the charge current flows in the reverse direction ($FM_{fixed} \rightarrow FM_{free}$) (Fig. 2b), the unpolarized electrons become polarized by the $FM_{free}$ layer and are then incident on the $FM_{fixed}$ layer. Similarly, the transverse spin component of this incident angular momentum tries to orient the magnetization of the $FM_{fixed}$ layer towards its direction by the spin torque. However, the exerted torque on the $FM_{fixed}$ is insufficient to switch the magnetization because of the high anisotropy of $\boldsymbol{M}_{fixed}$. Instead, the scattering process at the $FM_{fixed}$/NM interface causes a reflected electron flow, which is spin-polarized in the direction opposite to that of the $FM_{fixed}$ owing to the conservation of total angular momentum. This reflected electron flow with reverse spin angular momentum travels back to the $FM_{free}$ layer. The transverse spin component of the incident angular momentum due to the reflected electron is now absorbed by the $FM_{free}$ layer and exerts spin-torque on the layer to rotate $\boldsymbol{M}_{free}$ towards the reflected spins. Hence, an antiparallel configuration between the magnetization of $FM_{free}$ and $FM_{fixed}$ is now stabilized for the reverse current flow. Note that similar angular momentum reflection also happens in Fig. 2a, but the spin-torque is insufficiently exerted on the $FM_{fixed}$. Therefore, the STT can generate parallel or antiparallel spin configurations between the two FM layers. The $I$-d$V$/d$I$ characteristic of a typical STT nanopillar structure (e.g. Co/Cu/Co) is shown in Fig. 2c, where an asymmetric response can be seen [8]. Typically, the critical current to induce magnetic switching is approximately $10^{10}$ A $m^{-2}$. Both parallel and antiparallel spin switching with the current flow direction can be useful for read-write applications [8,58].

***2.2 Spin-orbit torque***

The SOT is an alternative approach with a higher spin-charge conversion efficiency than STT, reducing critical current densities for switching devices. Unlike STT using a vertical current, an in-plane charge current is utilized in SOT without the additional FM polarizer layer (Fig. 3a). A traditional SOT device consists of an FM/NM bilayer heterostructure. When an in-plane charge current passes through the NM layer or the FM/NM interface, SOT is developed at their interface because of the coupling between the electron spin and its orbital motion, *i.e.* spin-orbit coupling (SOC) [1]. As a consequence, non-equilibrium transverse spin currents are generated along the out-of-plane direction and accumulated at the interface, which is responsible for the spin-torque acting on the FM magnetization. Simultaneously, a spin current polarization, $p_s$, is generated perpendicular to the plane defined by both $j_c$ and $j_s$. Compared with the STT devices, the design of SOT devices is easier as the magnetization (in-plane and out-of-plane) of the FM layer can be controlled by a smaller in-plane write current. Another advantage of the SOT devices is that the orientation of the magnetic state can be identified by passing a small out-of-plane read current [19]. Therefore, SOT-based devices such as SOT-Magnetoresistive random-access memory (MRAM) and SOT- Magnetic tunnel junction (MTJ) are more effective for spintronics owing to their robust design, low power consumption, and faster speed than STT-based MRAM and MTJ devices [61].

Figure 3b shows a simple model of the generated spin-torque, wherein $\boldsymbol{H}_{\mathrm{eff}}$ is the effective magnetic field including of applied external, dipolar and anisotropy fields; $\boldsymbol{M}$ is the local magnetization. Assuming that there is no anisotropy, $\boldsymbol{H}_{\mathrm{eff}}$ contains the external magnetic field and magnetic dipolar field induced by the magnetic layers. Without electrical current, if local magnetization ($\boldsymbol{M}$) is tilted away from an effective magnetic field ($\boldsymbol{H}_{\mathrm{eff}}$), two types of torques appear, namely an effective field torque ($\boldsymbol{T}_{\mathrm{field}} = -\boldsymbol{M}\times\boldsymbol{H}_{\mathrm{eff}}$) and a damping torque ($\boldsymbol{T}_{\mathrm{damping}} \propto \boldsymbol{M}\times\mathrm{d}\boldsymbol{M}/\mathrm{d}t$) [62]. The former induces a precessional movement of $\boldsymbol{M}$, whereas the latter rotates $\boldsymbol{M}$ to align it along the $\boldsymbol{H}_{\mathrm{eff}}$. In the presence of an applied current, a new torque can be generated, which can be decomposed into two components, namely damping-like torque ($T_{\mathrm{DL}}$) and field-like torque ($T_{\mathrm{FL}}$) [58–60,62–65]. $\boldsymbol{T}_{\mathrm{DL}}$ acts either along or opposite to the direction of $\boldsymbol{T}_{\mathrm{damping}}$, depending on the direction of the current. As a consequence, it behaves as an additional damping or antidamping source. Notably, $\boldsymbol{T}_{\mathrm{DL}}$ is an important factor for the reversal of $\boldsymbol{M}$ towards its final equilibrium position. When $\boldsymbol{T}_{DL}$ is in the same direction as $\boldsymbol{T}_{damping}$, the spin-polarized current enhances the magnitude of the total effective damping, leading to the rapid dissipation of the $\boldsymbol{M}$ oscillation energy. Similarly, the direction of $\boldsymbol{T}_{FL}$ also depends on the direction of the current. Note that the spin-torque mechanisms (both damping-like and field-like torques) shares similar basic concepts for STT and SOT. The total magnetization in the final equilibrium state due to spin-torque (e.g. STT and SOT) is thus due to the combined effects of both $\boldsymbol{T}_{DL}$ and $\boldsymbol{T}_{FL}$ [65,66].

There are two key mechanisms for inducing SOT: spin Hall effect [66] and Rashba-Edelstein effect [67]. Both phenomena generate non-equilibrium spin accumulation at the FM/NM heterostructure interface and modify the magnetization dynamics of FM. The spin Hall effect mechanism represents a collection of SOC phenomena at the NM layer, in which unpolarized charge currents can produce transverse spin currents and vice versa (Fig. 3c). This effect was first proposed by Dyakonov and Perel in 1971 and more recently by Hirsch in 1999 [42,68]. The basic mechanism of spin Hall effect in FM is closely interlinked with the anomalous Hall effect (AHE). However, unlike FM materials, NM materials have equal number of spin-up and spin-down electrons in equilibrium without any charge imbalance. When an in-plane charge current is applied in NM materials, asymmetric spin-dependent scattering occurs because of the SOC. Consequently, spin-up and spin-down electrons are deflected in opposite directions, inducing a transverse spin current, $\boldsymbol{j}_s \propto \theta_{SH}(\boldsymbol{j}_c \times \boldsymbol{p}_s)$; where $\theta_{SH}$ is known as spin hall angle or SOT efficiency [69,70]. Such a deflection gives rise to a spin Hall voltage at the FM/NM interface. Moreover, both the sign and magnitude of $\theta_{SH}$ and the Hall voltage are indicative of the intrinsic characteristics of the NM such as the spin-current generation efficiency, SOC, and asymmetric carrier density.

In some FM/NM heterostructure systems, the interfacial SOC generated from both the crystal structure of the NM and broken crystal symmetry gives rise to the SOT mechanism, namely the Rashba-Edelstein effect (Fig. 3d) [19,67,71–73]. This symmetry breaking lifts the spin degeneracy in *k*-space and shifts the valence band maxima and/or conduction band minima from their symmetry points. The modification of the band structure by the Rashba splitting energy in *k*-space can have a strong impact on the NM charge carriers [61,73]. At the same time, an electric field also generates symmetry breaking in the NM. Consequently, the moving conduction electron experiences an effective force perpendicular to the direction of both the electric field and spin angular momentum. This force now couples with the spin angular momentum of the conduction electrons and acts on up- and down-spins along opposite directions. The resultant polarized spin accumulation at the FM/NM interface gives rise to SOT and magnetization switching. Detailed description on the microscopic origin of the spin Hall effect and Rashba-Edelstein effect can be found elsewhere [74–80]. It is worth noting that the functionalities of SOT devices can be tuned to meet the requirements of spin-based electronics to a large extent by adjusting the material properties with the help of impurity doping and structural modulation of the NMs, conventional device geometry design, or by modifying the injected current and output voltage response.

## 3. Probing the SOT

The SOT effect can be detected via magnetization orientation of the switchable FM layer under different excitation conditions (e.g. amplitude of the applied current or $H_{ext}$) [81]. To evaluate the efficiency of the SOT, the spin Hall angle ($\theta_{SH} = j_s/j_c$) should be extracted. Note that the current injection may increase the magnetothermal effect caused by the Joule heating, which can also affect the magnetization dynamics. Therefore, special care should be taken during the experiments. Keeping the SOT device inside a liquid nitrogen cryostat during the measurements is one possible solution to reduce the Joule heating [82]. Here, we will explain in detail three conventional techniques to investigate the SOT phenomena and efficiency: i) Hall measurements (e.g. anomalous Hall effect ($R_{xy}$)) [83], ii) spin-torque ferromagnetic resonance (ST-FMR) measurements using the anisotropic magnetoresistance ($R_{AMR}$) [10,84], and iii) optical measurements using the magneto optical Kerr effect (MOKE) [85,86].

### *3.1 Hall measurements*

A conventional Hall bar geometry can be used to characterize the SOT (Fig. 4a) [24,87]. The key concept is to detect the anomalous Hall effect signal (strongly dependent on magnetization of the FM layer) of the structure which can be modulated by applying different currents. To evaluate the SOT efficiency, the Hall resistance ($R_H$) on the *xy*-plane is investigated as functions of out-of-plane $H_{ext}$ at different amplitudes of DC currents (in-plane) (Fig. 4b). Here, the saturated hysteresis loops originate from the rapid reorientation of the magnetic domains along $H_{ext}$, which results in the effective field and $R_H$ reaches their maximum values at the saturation field. The measured field-dependent resistivity in this SOT heterostructure shows a hysteresis loop shift, which exhibits coercive field changes ($\Delta H_c$) at different values of currents. In this situation, $\theta_{SH}$ can be easily calculated by using $\theta_{SH} = \frac{4\pi e}{h} M \times t_{FM} \frac{H_{so}}{j_c}$, where $H_{so}$ is the out-of-plane spin-orbit field, $t_{FM}$ is the thickness of the FM layer, and $h$ is the Planck constant. One can consider a macrospin switching model to determine $H_{so}$ at different currents [82,88]. When the FM magnetization has only an in-plane component and no out-of-plane component at $H_c$, the induced $\boldsymbol{H}_{so}(\hat{z})$, which counters $\boldsymbol{H}_{ext}(-\hat{z})$ (inset of Fig. 4b), significantly affects $\Delta H_c$. Therefore, the measured $\Delta H_c$ is almost equal to $H_{so}$ according to micromagnetic simulations, which can be used to estimate the $\theta_{SH}$ in the Hall bar structure [82,88]. It is noted that the in-plane magnetic anisotropy contributes to the hysteresis loop if in-plane magnetic anisotropy or remanent magnetization along in-plane direction is not completely zero. Nevertheless, the in-plane magnetic anisotropy is normally quite small in many magnetic materials and can easily reorient magnetization with a small out-of-plane external magnetic field during SOT measurements. For example, the MnGa layer in BiSb/MnGa bilayer has a uniaxial easy axis along the z direction, and biaxial easy axes at $\varphi = \pm 45°$ from the z direction with no in-plane magnetic anisotropy [82].

To determine direction and amplitude of the torque, which cannot be evaluated by a simple Hall measurement with DC current, the harmonic Hall measurement technique is used by applying AC current under different directions of in-plane $H_{ext.}$ The total Hall voltage $V_H$ with AC current consists of the anomalous Hall effect and planar Hall effect (PHE) (Hall signal generated by an applied in-plane magnetic field) signals across the *xy*-plane in the Hall bar structure. It can be expressed as $V_H(I_{ac})=R_{AHE}\times I_{ac}\times\cos\theta + R_{PHE}\times I_{ac}\times\sin^2\theta\times\sin2\phi$, where $R_{PHE}$ is the Hall resistance induced by the planar Hall effect signal [21,87]. By measuring $V_H$ at $\phi = 0°$ and 45°, one can easily determine both the $R_{AHE}$ and $R_{PHE}$ at a fixed current. Because both $\theta$ and $\phi$ are functions of $I_{ac}(t)$, the Hall resistance can be written as $\frac{V_H(I_{ac})}{I_0} = R_H^f \times \sin(ft) + R_H^{2f}(I_{ac}) \times \cos(2ft)$, where $R_H^f$ and $R_H^{2f}$ are the first- and second-harmonic Hall resistances, respectively. In general, the observed $R_H^f$ $(= R_{AHE} \times \cos\theta + R_{PHE} \times \sin^2\theta \times \sin 2\phi)$ is almost the same as the DC-$R_H$ in a similar Hall-bar structure. However, $R_H^{2f}$ is preferably used to analyze the SOT by adopting the convenient approximation so that the rotating $H_{ext}$ is applied in-plane and $\boldsymbol{M}$ rotates with the constant in-plane $H_{ext}$ (> perpendicular anisotropy field, $H_A$) [21,87]. In this scenario, $H_{ext}$ and local magnetization point towards random directions with $\phi$. The change in $R_H^{2f}$ as a function of $\phi$ is measured at different constant $H_{ext}$ (Fig. 4c). These curves individually reflect the sum of $R_{DL}$ (damping-like SOT component), $R_{FL}$ (field-like SOT component), and $R_{Oe}$ (Oersted field SOT component). It has been established that $R_H^{2f}$ contains both cos$\phi$ and cos(3$\phi$) terms according to the relation $R_H^{2f} = \cos\phi\left(R_{DL} + \frac{R_{FL}+R_{Oe}}{2}\right) + \cos 3\phi\left(\frac{R_{FL}+R_{Oe}}{2}\right)$ [39,89]. A theoretical simulation of the experimental data provides the strengths of all the active SOT components. Additionally, analysis of the dependence of $R_H^{2f}$ on $H_{ext}$ needs to be performed to separate all the SOT components to calculate the acting SOT fields (e.g. damping-like field ($H_{DL}$), field-like field ($H_{FL}$), and Oersted field ($H_{Oe}$)). In this case, $R_H^{2f}$ can be expressed as $R_H^{2f} = \cos\phi\left(R_{AHE}\frac{H_{DL}}{H_{ext}-H_A} + R_{PHE}\frac{H_{FL}+H_{Oe}}{H_{ext}}\right) + \cos 3\phi\left(R_{PHE}\frac{H_{FL}+H_{Oe}}{H_{ext}}\right)$. By comparing the coefficients of cos$\phi$ and cos(3$\phi$) from the two expressions of $R_H^{2f}$, the effective SOT fields can be easily estimated. It should be noted that the sets of $R_H^{2f}$ vs. $\phi$ or $H_{ext}$ measurements required depend on the number of unknown components present. Accordingly, the SOT efficiency can be estimated by using $T_{DL/FL} = \frac{4\pi e}{h} M_{FM} t_{FM} \frac{H_{DL/FL}}{j_{ac}}$ [21,87].

### *3.2 Spin-torque ferromagnetic resonance*

The ST-FMR relying on the ferromagnetic resonance (FMR) technique is another approach for investigating the magnetization dynamics of FM materials as well as SOT efficiency with the help of microwave-frequency charge current ($j_{rf}$) [10,84].The sensitivity of the ST-FMR method is high enough to detect the output SOT signals from microsized or even nanosized spintronic devices. The method was first utilized to demonstrate the SOT efficiency in a Pt/NiFe heterostructure planar spin Hall bar device in 2011 [84].

Figure 4d shows a schematic of the ST-FMR measurement setup and its results. An oscillatory $j_{rf}$ injected into the SOT heterostructure can generate an oscillatory current that induces non-equilibrium spin accumulation at the interface due to the SOC. The accumulated spins diffuse into the adjacent FM layer and exert oscillatory SOTs ($T_{DL}$ and/or $T_{FL}$) on the local magnetization. As a result, the local magnetization precesses around the $H_{eff}$, leading to an $R_{AMR}$ in the SOT device. The bias-tee is used in this setup to apply a $j_{rf}$ and simultaneously measure the output voltage. The voltage drop in the heterostructure device is detected by either a DC voltmeter or a lock-in amplifier. The DC and AC techniques can detect a signal resolution on the order of microvolts and 10 nanovolts, respectively [10,90,91]. The detected output voltage depends on the amplitude of $j_{rf}$, the $R_{AMR}$ of the device, ϕ (azimuthal-angle of $\boldsymbol{M}$), $\theta_c$ (cone angle of the $\boldsymbol{M}$), and φ (resonance phase between $T_{DL}$ and/or $T_{FL}$ and $\boldsymbol{M}$) [10]. The output voltage is a combination of a symmetric function and an asymmetric Lorentzian function, *i.e.* $V_{mix}=V_sF_s+V_aF_a$, where 's' and 'a' stand for the symmetric and asymmetric parts of the signal, respectively (Fig. 4e) [84]. The symmetric amplitude ($V_s$) of the Lorentzian function is related to the $T_{DL}$ by the relation $V_s \propto j_s h/(4\pi e\mu_0 M\times t_{FM})$, while the asymmetric amplitude ($V_a$) is directly connected to the oscillatory torques on $\boldsymbol{M}$ (i.e. $T_{FL}$+ rf-driven Oersted field torque) by the formula $V_a \propto H_{Oe}\times[1+(4\pi M_{eff}/H_{ext})]^{1/2}$, where $M_{eff}$ is the effective magnetization of the FM layer [10,84]. Furthermore, the SOT efficiency can be easily estimated from the ratio of the Lorentzian components (dotted curve), *i.e.* $\theta_{SH}=(V_s/V_a)(2\pi e\mu_0 M_s t_{FM} t_{NM}/h)[1+(4\pi M_{eff}/H_{ext})]^{1/2}$ [10,33,92]. Currently, this technique is extensively used to explore SOT in various magnetic heterostructures such as MTJs [93], spin valves [91], and magnetic semiconductors [94].

### *3.3 Magneto-optical Kerr effect*

Polarized light interacts with the magnetic order of materials. Magneto-optic techniques helps to determine the magnetic structure or the optically active magnetization [85,86], and domain structures or spin density of states [95,96]. However, both current-induced in-plane and out-of-plane magnetization dynamics and SOT efficiency of magnetic heterostructures can be quantified by analysing the light polarization rotation (either reflected or transmitted light) due to the MOKE in either harmonic [96–98] or pump-probe techniques [99]. When linearly-polarized laser light is incident on the surface of a magnet, the polarizations of the reflected and transmitted light change according to the magnetization orientation of the magnet [97]. This polarization change in SOT devices is directly affected by the magnetization dynamics of the FM layer due to the spin-torque. Here, we will discuss polar MOKE measurements where the laser light is normally incident on the sample surface and the polarization of the reflected laser is sensitive only to the out-of-plane magnetization of the FM.

Figure 5a shows the schematic of a typical polar MOKE setup. Orientation of the magnetization at the laser spot is modulated by the SOT when an in-plane $j_{ac}$ is injected into the NM layer. Each component of the switched magnetization can be detected through the Kerr rotation angle ($\theta_K$), which can be recorded by a lock-in amplifier coherent with the frequency of $j_{ac}$ [96,97]. Here, the balanced detector is used to analyse the voltage difference ($\Delta V$) between two separate linearly polarized beams with orthogonal polarization of the reflected laser signal after the laser passes through a Wollaston prism (beam splitter). The observed Kerr signal exhibits a sharp change near $H_{ext} \rightarrow H_c$ due to magnetization reversal while the other portion of the signal depends weakly on $H_{ext}$ [97,100]. Changing the current modulates the magnetization of the FM layer, which can be detected from the MOKE signal. The dynamics of the SOT can be detected by adding another lock-in amplifier (figure not shown) [97].

In addition, time-resolved (TR) MOKE can probe the SOT-induced magnetization switching and domain wall motion, as shown in the schematic of a TR MOKE setup in Fig. 5b [101]. The pump-probe stroboscopic technique is useful for investigating regularly repeating motions in the magnetization dynamics due to SOT. The electrical pulse (pump) plays a role in generating the current flow for SOT. The MOKE signals in the reflected picosecond laser beam (probe) from the FM layer are used for detecting the corresponding magnetization dynamics. The electrical pulse is synchronized to the picosecond laser beam with a well-defined delay time, which determines the temporal resolution. Furthermore, it is expected that the switching mechanisms of the SOT, including $T_{FL}$ and $T_{DL}$, would be more obvious from the temporal analysis of the SOT-driven magnetization dynamics. It is important to note that the magnetization of FM layer returns to its initial equilibrium state between the pump pulses of current. To date, there are only a few reports on probing SOT by the optical technique for various systems like metallic ultra-thin Ta/CoFeB/MgO [97,102], $Y_3Fe_5O_{12}$/Pt [97] and $Ni_{80}Fe_{20}$/Pt [103].

## 4. Traditional materials for SOT

The SOT device can be improved by choosing efficient and appropriate materials for both the torque and FM layers [81,104]. The torque layers are NM with high SOC and/or induced structural inversion asymmetry at the interface of the FM layer that can generate strong current-driven torque via the Spin Hall or the Rashba-Edelstein effects. A few antiferromagnetic (AFM) materials have also been utilized instead of NMs owing to their own anomalous Hall effect [105,106]. The key characteristic for the SOT efficiency of the torque layer is $\theta_{SH}$, which represents the spin-charge conversion efficiency. A high $\theta_{SH}$ is required for an energy efficient SOT device. Furthermore, a high conductivity ($\sigma$) in the torque layer is as important as $\theta_{SH}$. It ensures that the current flows through the torque layer instead of the FM layer at FM/NM heterojunction. Otherwise, the applied $j_c$ cannot generate $j_s$ in the FM/NM (when $\sigma_{FM} > \sigma_{NM}$) heterostructure system because the $j_c$ is shunted through the FM layer from the NM layer and then passes through the FM layer without producing any $j_s$ in the NM layer [107]. The power consumption ($P \propto 1/(\sigma\times\theta_{SH}^2)$) for switching a unit magnetic volume in the FM, is also an essential parameter in choosing the best torque materials for SOT devices with low power consumption [108]. Besides these key parameters, the spin Hall conductivity ($\sigma_{SH}$), which reflects the magnitude of the spin accumulation at the interface, is also an important figure of merit in realistic SOT applications. Obviously, the operating temperature of the device should be near room temperature (RT) for practical applications. Table 1 summarizes $\theta_{SH}$, $\sigma$, $\sigma_{SH}$, $P$, and working temperature of various torque materials. FMs for SOT must have low coercive fields (e.g. a few tens of Oersted) for easy magnetization switching and lower conductivities than the torque layers for the generation of $j_s$ [107].

### *4.1 Nonmagnetic materials for SOT*

Among the various NMs, heavy element-based NM metals (5*d*-transition metals such as Ta, W, and Pt) are more attractive because of their stronger SOC compared with light element-based NM metals (Ti and Cu) [22–24]. Heavy metals such as Pt [23,84,109], β-W [110], β-Ta [107,111], Hf [112], and W(O) [110] are widely investigated for SOT applications. Among these NM metals, only W in both β-phase and oxidized state shows large SOT efficiencies. Other NM metals exhibit low SOT efficiency because of the opposite signs of their bulk and interfacial SOC effects [22,111,112]. Besides these single NM metals, bilayer NM systems such as W/Hf [113], Pt/Hf [114], Pt/Ta [115], Pt/W [116], and Pt/Ti [117], as well as Rashba interface materials such as $LaALO_3/SrTiO_3$ [11,118], and Bi/Ag [106], are also widely investigated. The corresponding studies confirm that the SOT efficiency can be amplified by combining two-material structures [119,120]. A significantly higher SOT efficiency in a 2D electron gas formed at the $LaAlO_3/SrTiO_3$ interface because of the direct Rashba-SOC that has been observed at RT [11,36,118]. Interestingly, the SOC of these NM metals can be significantly enhanced through natural oxidation, such as in Cu(O) [10,121] and/or impurity doping, such as in $Cu_{1-x}Pt_x$ [122], $Cu_{99.5}Bi_{0.5}$ [123] and $Cu_{97}Ir_3$ [124]. For example, the alloy $Cu_{99.5}Bi_{0.5}$ shows a larger $\theta_{SH}$ than the heavy metals Pt and Ta [84,107,123]. Recently, a new type of NM/FM/NM multilayer (e.g. Ta/Au/FM/Au/$AlO_x$) has been found to exhibit very strong $T_{DL}$ owing to the generation of the planar Hall current in the FM layer when one of the NM/FM interfaces in the NM/FM/NM multilayer behaves like a good spin-transfer source and the other interface like a spin sink [125].

### *4.2 Antiferromagnets for SOT*

Promising SOT results (Table 1) have recently been observed when non-collinear AFMs (such as PtMn and IrMn) are used instead of NMs because of the self spin Hall or anomalous Hall effects in the AFMs [105,106]. This approach uses an antiferromagnet as the torque layer because the AFM can exert an internal exchange bias field on the adjacent FM layer without the assistance of any external field [126,127]. As a result, it can easily switch the magnetization of FM layer along the out-of-plane direction via the SOT that emerges from the direct spin Hall effect in these torque layers. A collinear IrMn AFM structure has exhibited a comparatively large SOT efficiency of approximately 0.6 compared to the heavy metal β-W ($\theta_{SH} \sim 0.33$) [102] and polycrystalline IrMn ($\theta_{SH} \sim 0.08$) [106]. Magnetization reversal in AFM/FM heterostructure systems through SOT without $H_{ext}$ can be a future core technology for read-write and memristor applications [53,59,94,105].

## 5. Topological materials for SOT

As a part of the exploration for new materials that exhibit significantly large current-driven toque, topological insulators (TI) are found to be potential candidates for future SOT-MRAM [30,128]. TIs are quantum materials that have insulating bulk states and metallic surface states [1]. TIs were extensively studied in the early 1980s in research on the quantum Hall effect, which originates from the non-trivial topology of the two-dimensional electron wavefunctions under a strong magnetic field [25]. It was later predicted that the quantum Hall effect can be realized without external magnetic fields (quantum- anomalous Hall effect), based purely on topology arguments [26]. It is worth noting that the quantum spin Hall effect is another important aspect of topological insulators for the low-dissipation transport [129–132]. The quantum spin Hall effect is quantization of the spin Hall effect, which is analogue to the quantum Hall effect (quantization of Hall effect). Quantum spin Hall effect can be considered as the spin version of quantum Hall effect, wherein the carrier transport at the edge is spin-polarized [131,132].

The first 2D TI with spin-polarized edge states was predicted in a heterostructure consisting of a HgTe quantum well sandwiched by two (HgCd)Te barriers [133] and was immediately confirmed experimentally [134]. Three-dimensional (3D) TIs were proposed [135,136] and first confirmed in BiSb [137], followed by several Bi-based chalcogenides, such as $Bi_2Se_3$, $Bi_2Te_3$, and $(BiSb)_2Te_3$ [138–141]. TIs have many different characteristics including quantum Hall states that make them attractive for RT spintronic applications: i) Existence of the one (two)-dimensional edge (surface) states of TIs is ensured by the non-trivial topologies of their band structures and can emerge without the application of large external magnetic fields, ii) the surface states have Dirac-like band dispersions, which promise a large intrinsic spin Hall effect due to Berry phase mechanism, and iii) a unique spin-momentum locking feature of the topological surface states prioritizes pure $j_s$ generation in the direction perpendicular to the film plane [27–29].

Recent experiments have demonstrated large $\theta_{SH}$ at RT in $Bi_{0.9}Sb_{0.1}/Mn_{0.5}Ga_{0.55}$ heterostructure because of both surface and bulk spin Hall effect [82]. The large $\theta_{SH}$ was estimated from Hall bar measurements of a $Bi_{0.9}Sb_{0.1}$ (10 nm)/$Mn_{0.5}Ga_{0.55}$ (3 nm) heterostructure with a small tilted magnetization and the change in Hall resistance at different $j_c$ and $\theta = 2°$ (angle between $H_{ext}$ and $z$-axis) (Fig. 6a). The changes in $\Delta H_c$ at different given values of $j_c$ in $R_H$ vs. $H_{ext}$ plot are exactly equal to $H_{so}$, which counters $H_{ext}$ and is aligned opposite to $H_{ext}$ (as mentioned in the inset of the $R_H$ vs. $H_{ext}$ plot). The ratio of $H_{so}/j_c$ ($\sim\Delta H_c/j_c$) is used to calculate $\theta_{SH}$ as discussed in probing section. The estimated $\theta_{SH}$ at RT is 52 [82]. In addition to $Bi_{0.9}Sb_{0.1}$, strong spin Hall effect has been observed in various TIs (Table 1) such as $Bi_2Se_3$ ($\theta_{SH} = 2\text{-}3.5$ at RT) [30], $Bi_xSe_{1-x}$ ($\theta_{SH} = 18.8$ at RT) [142], $(Bi_{0.5}Sb_{0.5})_2Te_3$ ($\theta_{SH} = 25$ at 200 K) [31], and $(Bi_{0.5}Sb_{0.5})_2Te_3$ ($\theta_{SH} = 140\text{-}410$ at 1.9 K) [128]. SOT magnetization switching with ultralow current densities has been demonstrated at RT in $Bi_2Se_3$/CoTb, $(BiSb)_2Te_3$/CoTb [108], $Bi_2Se_3$/NiFe [33], $Bi_xSe_{1-x}$/Ta/CoFeB/Gd/CoFeB [142], and BiSb/MnGa [82].

Although giant spin Hall angles and low SOT switching current densities have been confirmed in various TI/FM bilayers, the origin of the giant spin Hall effect in TIs is still unclear because the fact that the current may flow in both the surface and bulk of TIs. To definitively determine the origin of the giant spin Hall effect in TIs, the spin Hall effect in $(Bi_{1-x}Sb_x)_2Te_3$/Ti/CoFeB was investigated at various Sb compositions with different Fermi levels (Fig. 6b) [32]. When the Sb compositions are about 85% and 93%, the Fermi levels are in the band gap of $(Bi_{1-x}Sb_x)_2Te_3$ and approach the Dirac point [32,141]. Near the Dirac point, the switching current density is minimized and the SOT-induced effective field maximized. This unambiguously demonstrates the surface state origin of the giant spin Hall effect. Furthermore, SOT efficiency depends on the layer thickness of TIs. The TI-based SOT device shows large $\theta_{SH}$ when the thickness of TI layer goes to a monolayer limit, resulting from the formation of topological surface states and reduction of 2D electron gas states as well as bulk states (e.g. $Bi_2Se_3$/Py heterostructure) [33]. Recently, gate-tunable $\theta_{SH}$ is observed in Cr impurity- doped TI-based SOT device [34].

For realistic applications in SOT-MRAM, both $\theta_{SH}$ and $\sigma$ are equally important for reducing the writing power consumption. With respect to this, it was observed that the narrow bandgap TI BiSb is the most promising material because it satisfies both the above criteria (high $\sigma \sim 2.5\times10^3\ \Omega^{-1}\mathrm{cm}^{-1}$ and large $\theta_{SH} \sim 52$) [82,143]. In addition, a strong interfacial Dzyaloshinskii-Moriya interaction in BiSb/MnGa [144] and a giant unidirectional spin Hall magnetoresistance of 1.1%, which is three orders of magnitude larger than those in metallic bilayers, have been observed in a BiSb/GaMnAs bilayer [145]. Thus, BiSb has been considered as one of the best SOT materials for realistic SOT devices.

## 6. 2D materials for SOT

The first key expectation from 2D magnetic materials is the low density of their spin-polarized carriers, which can be tuned by different approaches (e.g. electrostatic gating [47], and external electric field [47,49,50]) rather than the traditional magnetic field. Interestingly, the weak interaction between two adjacent layers introduces antiferro-ferro switching by an external electric or magnetic field [50]. Another importance of 2D magnetic materials is their possibilities to shed light on the proximity effect between two different quantum materials that occur at the interfaces of two bulk or thin films, especially in magnetic and superconducting materials. Many exotic phenomena appear across a wide range of condensed matter physics such as magnetic exchange bias, enhancement of superconducting transition temperature, and superspin-current generation [73,146]. However, such physics at the interfaces are hidden from surface analysis tools. This problem can be resolved if 2D magnetic materials are utilized. Furthermore, the proximity effects, which normally appear at a very thin layer at heterostructure interfaces, occur throughout the entire 2D thin layers, strengthening the effect at the interface. The SOT can be a representative example of this advantage of 2D magnetic materials. Here, we will briefly discuss about the 2D layered materials for SOT devices as a torque, FM and both layers.

### ***6.1*** *2D materials as torque layers*

Transition metal dichalcogenides (TMDs) consisting of one heavy transition metal and two chalcogen atoms exhibit strong SOC band structures, which can generate current-driven out-of-plane spin polarization [147–149]. Such a strong SOC can induce a large spin Hall effect as well as Rashba-Edelstein effect at the interface with an adjacent material. SOT has been demonstrated in recent experiments in which 2D TMDs were used as the spin source layers owing to their tunable structural symmetry [38,39,102], atomically flat surfaces, broken inversion symmetry even in monolayer range [150,151], gate-modulated SOC [152,153], adjustable out-of-plane magnetic anisotropy [82], and tunable self- anomalous Hall and spin Hall effects [153,154]. The 2D Weyl semimetal $WTe_2$ is extensively used in SOT devices as a torque layer owing to its topological nature governed by its reduced crystal symmetry [38,102,155102]. Interestingly, while almost-thick SOT devices reveal in-plane $T_{DL}$, out-of-plane $T_{DL}$ and magnetization switching have been reported in a $WTe_2$ (1.8-15 nm)/Py (6 nm) heterostructure at RT [38]. Figure 7a shows a ST-FMR measurement setup and the results in a $WTe_2$/Py heterostructure. The result represents the ST-FMR signals at RT and a frequency of 9 GHz with two different directions of magnetization rotated 180° with respect to each other (e.g. $\phi_1 = 40°$ and $\phi_2 = 220°$). The signals at these two 180° rotated directions will be symmetrical if the system has two-fold rotational symmetry. However, the voltage signal in the $WTe_2$/Py bilayer is completely asymmetric (e.g. $V_{mix}(\phi = 40°) \neq -V_{mix}(\phi = 220°)$), indicating a low-symmetry SOT current induced by the low-symmetry crystal $WTe_2$. Furthermore, it has been confirmed that the SOT efficiency in the $WTe_2$/FM heterostructure can be tuned with crystal symmetry [38] as well as small critical current [48]. However, a comparable $\theta_{SH}$ of approximately 0.013 with Pt/Py heterostructure [23] was estimated in $WTe_2$/Py bilayer from the variation of $V_{mix}$ with $H_{ext}$ using the $V_s/V_a$ ratio method (details in probing section) [38]. Also, the SOT efficiency of this device structure weakly depends on the thickness of $WTe_2$ layer but strongly depends on the different FM layer [155]. Owing to the out-of-plane $T_{DL}$ and strong SOC, this 2D material has the potential to replace existing 3D torque layers for SOT devices.

A high SOT efficiency of approximately 0.14, which can exert both $T_{DL}$ and $T_{FL}$ on the FM layers (e.g. CoFeB, Py at RT [39,156]), has also been observed in the layered 2H-$MoS_2$. These two torques are strong enough to excite FMR in the FM layer [156]. However, the observed $T_{DL}$ is much larger than the $T_{FL}$ from ST-FMR measurements in huge difference between symmetric and asymmetric ST-FMR peaks. Recently, the layered TMD $WSe_2$ was used as a torque layer because of its strong Rashba-Edelstein-type SOC owing to low crystal symmetry [39,157]. This material can generate out-of-plane $T_{DL}$ with Py at RT. It was suggested that monolayer $WSe_2$ in a Ta/$WSe_2$ heterostructure has the ability to enhance the SOT efficiency of Ta [157]. The low crystal symmetry β-$MoTe_2$ shows out-of-plane $T_{DL}$ and layer dependent SOT efficiency with a low conductivity as that of $WTe_2$ [158]. Experimentally, A fully metallic and highly symmetric TMD $NbSe_2$ exhibits out-of-plane Oersted torque as well as in-plane $T_{DL}$ with comperatively high conductivity ~$10^3$ $\Omega^{-1}cm^{-1}$ (comparable with $WTe_2$ [38]) [159], which can be slightly tuned by the thickness modulation and crystal strain. Therefore, both crystal symmetry breaking and crystal strain are the possible origin of SOT in 2D TMDs. Recently, a type-II Dirac semimetal $PtTe_2$ shows high SOT strength ~ 0.15 with a high conductivity ~$10^4$ $\Omega^{-1}cm^{-1}$ with a giant spin Hall effect, as the largest value compared to existing TMDs [39]. This TMD has also thickness-dependent SOT strength due to topological surface state like spin texture, suggesting future energy efficient and low-power SOT material.

***6.2*** *2D materials as FM layers*

As mentioned above, the switching efficiency depends on the magnitude ratio of the spin current to the total magnetization of the free layer, i.e. $j_s/(M_{free}\times t_{free})$ or $(j_c\times p_c)/(M_{free}\times t_{free})$ [60]. Furthermore, the torque occurs at the interface only within the spin diffusion length, which is usually quite short [160,161]. Therefore, an FM layer consisting of one or up to a few monolayers can improve the efficiency of SOT devices. Obviously, the introduction of 2D magnetic materials is the ultimate choice.

2D vdW FMs are emerging as an interesting research field owing to their layer-dependent properties and gate tunable Curie temperature ($T_c$) [162–164]. 2D vdW FMs also exhibit clean interfaces and strong out-of-plane magnetic anisotropy even at monolayer scales without any dead layers when such 2D FMs are stacked with different torque layers [11,38,165]. Among the various 2D FMs (e.g. $VSe_2$ [166], $Cr_2Ge_2Te_6$ [167,168], and $CrI_3$ [136]), $Fe_3GeTe_2$ has attracted considerable attention owing to its high $T_c$ (approximately 230 K for bulk crystals and 130 K for single atomic layers), out-of-plane magnetic anisotropy (a significant advantage for generating SOT), low coercive field ($H_c$ ~ 0.65 kOe), and a gate tunable $T_c$ of up to RT [162–164]. Figure 7b illustrates the harmonic Hall measurements and results in a $Fe_3GeTe_2$ (15–23 nm)/Pt (5 nm) heterostructure [11]. The second-harmonic Hall signals ($R_H^{2f}$) with rotating and constant magnetic fields (≥ anisotropy field of $Fe_3GeTe_2$) at the AC current amplitude of 2.4 mA are shown here for the $Fe_3GeTe_2$ (23 nm)/Pt (5 nm) heterostructure. Traditionally, $R_H^{2f}$ as a function of the AC current is a mixture of both cos ϕ and cos 3ϕ terms [11,89,165]. Interestingly, the observed $R_H^{2f}$ is well-fitted with only cos ϕ terms, indicating a small contribution by the field-like torque and the effect of the Oersted field [11]. This confirms that only $T_{DL}$ dominates the torque for switching the magnetization of $Fe_3GeTe_2$ because the cos(3ϕ) term is zero. A very large $T_{DL}$ (approximately 0.14 at $j_c$ ~ $2.5\times10^7$ A/cm$^2$ and 180 K) without any assistive of $H_{ext}$ compared with those in $Tm_3Fe_5O_{12}$/W ($T_{DL}$ ~ 0.02 at $j_c$ ~ $6 \times 10^6$ A/cm$^2$ and RT) [169], $Tm_3Fe_5O_{12}$/Pt ($T_{DL}$ ~ 0.014 at $j_c$ ~ $1.8 \times 10^7$ A/cm$^2$ and RT) [148,169], and $WTe_2$/Py ($T_{DL}$ ~ 0.013 at RT) [38] has been observed from $R_H^{2f}$ analysis [11]. The large $T_{DL}$ is attributed to the local domain wall motion in the $Fe_3GeTe_2$/Pt heterostructure.

Another significant result was observed in layered $Cr_2Ge_2Te_6$ 2D FM. The material has received immense scientific interest because of its semiconducting nature, gate-tunable Fermi level, and magnetic properties induced by the anomalous Hall effect, which is the source of spin Hall effect with different torque layers (e.g. Pt, Ta) [11,170]. It was also observed that the required $j_c$ of approximately $5\times10^5$ A/cm$^2$ for switching out-of-plane magnetization in a $Cr_2Ge_2Te_6$/Ta heterostructure is approximately two orders of magnitude smaller than that in a $Fe_3GeTe_2$/Pt heterostructure [11,170,171]. Recent observations of gate-tunable magnetic domains at RT [45] suggest that V-doped $WSe_2$ can be a future FM layer material for SOT devices owing to its long-range FM ordering along with semiconducting nature near RT.

***6.3** 2D materials for both torque and FM layers*

To date, all SOT experiments have been performed using bulk-NM/bulk-FM and/or bulk-FM/2D-NM, and/or 2D-FM/bulk-NM heterostructures. A combination of 2D materials may bring benefits to both the torque and magnetic layers. These heterostructures may also provide smooth and transparent interfaces so that comparatively small switching currents are required for magnetization reversal.

With respect to the use of 2D TMDs for both the torque and FM layers, the magnetization switching of $CrI_3$ (FM insulator) driven by SOT from $TaSe_2$ (NM metal) has been theoretically predicted in a bilayer-$CrI_3$/monolayer-$TaSe_2$ heterostructure (Fig. 7c) [118]. $CrI_3$ is one of the best choices for the 2D FM, as it exhibits layer-dependent magnetism [172]. For example, bilayer $CrI_3$ shows AFM ordering, while its monolayer behaves as a FM [172]. The weak exchange interaction between the layers of $CrI_3$ can be tuned from the AFM ground state to an FM state by using external magnetic or electric fields. It is therefore possible to use the spin-torque to manipulate this antiferro-ferro coupling switching. Meanwhile, $TaSe_2$ shows a huge broken crystal symmetry-driven SOC [148]. In this $CrI_3$/$TaSe_2$ heterostructure, passing an unpolarized charge current through the monolayer $TaSe_2$ generates an interfacial SOT that is strong enough to switch the magnetization of the first $CrI_3$ monolayer next to the $TaSe_2$ (the dynamics of magnetization switching are shown by the sphere for both the layers in the upper panel of Fig. 7c). As a result, the antiferromagnetically-coupled state of the bilayer $CrI_3$ is changed to the ferromagnetically-coupled state. The microscopic physical mechanism of this conventional SOT is described at the lower panel of Fig. 7c using the results of a first-principles quantum transport calculation. The torque acting on $CrI_3$ can only manipulate the magnetization of the bottom monolayer, which is the nearest neighbor to $TaSe_2$. The magnetization of the top $CrI_3$ monolayer away from the $TaSe_2$ remains unchanged because the spin density is zero in this layer (lower panel of Fig. 7c). The switching dynamics of the first $CrI_3$ monolayer can be verified by measuring the out-of-plane tunneling magnetoresistance and second-harmonic Hall responses experimentally, as suggested in Ref. [11].

## 7. Challenges and opportunities for 2D SOTs

For a decade from the discovery of SOT phenomenon, a lot of achievements in development of the SOT device have been made. The most important success of SOT device is to reduce the critical switching current down to ~ $10^5$ $Acm^{-2}$ by using topological materials and 2D materials [82,102102], which is totally adaptable for industrial requirements for memory applications. Nevertheless, an appropriate process to fabricate devices involving topological and 2D materials is still under development. Furthermore, the target of reduction of critical currents moves so fast that the investigation of underline mechanisms is overlooked. Therefore, the next step of the SOT research will focus more in growth techniques for materials and exploration of physical mechanisms. 2D vdW materials will be good candidates for such purposes. A search for materials with better performance will also be a mainstream in next decades. Together with memory applications, SOT devices showed possibilities for other important novel devices such as nano-oscillators [173,174], p-bit for probabilistic computing [175] for future neuron network hardware. For the last but not least, it is worth noting that the SOT-based devices use electrical current as the main driving force, which generates unavoidable energy loss due to joule heating. New multifunctional spintronic devices using gate bias or combining electrical current with gate biased should be focused in the near future [176].

### *7.1 2D topological insulators*

Most of the TI thin films studied so far were epitaxially grown on dedicated III-V semiconductor substrates by molecular beam epitaxy (MBE), which is not feasible for mass-production. Therefore, it is essential to investigate the performance of non-epitaxial TI thin films deposited on silicon substrates by an industry-friendly technique, such as sputtering and/or chemical vapor deposition and/or pulse laser deposition technique. A recent attempt to investigate the performance of sputtered $Bi_xSe_{1-x}$ TI thin films shows a promisingly large $\theta_{SH} \sim 18.6$ but a low $\sigma \sim 7.8\times10^1\ \Omega^{-1}cm^{-1}$ due to the poor crystal quality [142]. On the other hand, it has been demonstrated that high-quality BiSb thin films similar to those grown by MBE can be fabricated by sputtering deposition on sapphire substrates [177]. Further research will be needed to produce BiSb thin films with high crystal quality and spin Hall performance on Si substrates by sputtering deposition for ultralow-power SOT-MRAM.

### *7.2 New 2D magnetic materials*

Although a library of 2D magnetic materials has been established, all of these materials have quite low $T_c$ and high $H_c$. The requirements of stable magnetization with perpendicular magnetic anisotropy and suitable $H_c$ at RT are not met yet. Exploration of new 2D magnetic materials is still a challenging task. Beyond the conventional working principle, the weak interlayer magnetic interaction in 2D vdW materials provides opportunities for SOT devices with new designs, such as SOT-induced switching between interlayer magnetic states [118,172]. Recently noticed RT magnetism in monolayer $WSe_2$ semiconductor via vanadium dopant with reasonable $H_c$ can help to design new SOT devices for realistic and RT applications [45].

***7.3 2D magnetic semiconductors***

To prevent the applied current from flowing through the FM layers, one option is to use 2D FM semiconductors, which have high resistivities as the conventional 3D FM layers. The challenge is the limited availability of RT ferromagnetic semiconductors. The answer could be found in 2D diluted magnetic semiconducting TMDs [45,46]. This research field is under investigation. It is also important to note that the magnetic ordering and thermal fluctuation in SOT devices using 2D magnetic materials can be manipulated artificially by reducing layer thickness toward the monolayer limit [149].

***7.4 Heterostructure growth of 2D materials***

Similar to other spin devices, SOT devices require clean interfaces between the layers. Nevertheless, atomic-thick 2D materials are quite sensitive to the ambient environment. The incorporation of 2D materials by existing 3D fabrication techniques (e.g. sputtering, thermal/e-beam evaporation) without exposing the 2D materials to the ambient environment is still challenging. The *in-situ* growth of heterostructures could be a solution to minimize residues and environmental effects. SOT devices based on only 2D materials require vdW layer-by-layer growth, which is still premature at the current stage [35,178].

***7.5 Enhancement of conductivity of 2D TMDs***

2D semiconducting TMDs that exhibit high SOC at their valence bands can be used for torque layers. Their resistivities are often too high for incorporation with FM metals in SOT devices because all the current passes through the metallic layers instead of the TMD layers, which reduces the SOT efficiency [107]. Heavily doped TMDs may be a solution to this issue. Several candidates for p-doped TMDs such as V, Nb, and Ta have been demonstrated [45,179]. Shifting the Fermi level down to the valence band may enhance the spin Hall effect of the

TMDs.

### *7.6 Additional out-of-plane SOT*

The modern society magnetic memory technology needs out-of-plane magnetic anisotropy or perpendicular magnetization controlled by current or electric field instead of an external magnetic field. Magnetic switching devices with a perpendicular magnetization (e.g. perpendicular magnetic anisotropy) is strongly required to achieve large densities and thermal stability. However, almost-3D SOT devices exhibit high efficiencies only for in-plane torque on magnetic layers with in-plane magnetic anisotropy. Several approaches were demonstrated to couple the in-plane spin-torque with perpendicular magnetization by breaking the symmetry of the device structure or torque symmetry such as tilted magnetic anisotropy [180,181], interlayer exchange bias (e.g. using antiferromagnet/ferromagnet/oxides heterostructure) [105,106,182–184]. Incorporation of an additional in-plane ferromagnetic layer into the NM/FM (out-of-plane) junction can induce the vertical torque by changing the magnetization of the bottom layer [56].

Nevertheless, an out-of-plane spin-torque is required to high-efficiency perpendicular magnetization switching. The main approach in this research stream is to search materials with low symmetry point groups for the bilayer interface [185–187]. Recently, an $L1_1$-ordered (fcc staking order along [111] direction) CuPt (NM metals)/CoPt (FM) bilayer exhibits field-free out-of-plane magnetization switching of CoPt layer. The actual reason for such field-free effect is the low-symmetry point group (*3m*1) at the NM-metal/FM interface [185]. Furthermore, the incorporation of 2D materials into the SOT devices (e.g. $WTe_2$ for the torque layer [38,102,155102], $MoS_2$ as a substrate for Pt [188]) can also generate the additional out-of-plane torque exerted perpendicularly to the magnetic layers along with in-plane torques, giving rise to more efficient magnetization manipulation in SOT devices with perpendicular magnetic anisotropy. Understanding the generation mechanism of the out-of-plane spin torque is also an interesting subject.

### *7.7 Mechanism of SOT*

The spin Hall and Rashba-Edelstein band-type structures are the two accepted mechanisms for SOT. However, experimental evidence indicates unknown origins of SOT using harmonic study of the anomalous Hall effect and planar Hall effect in $AlO_x$/Co/Pt and MgO/CoFeB/Ta devices [87]. It was confirmed by space and time inversion symmetry arguments that SOT devices consisting of NM/FM heterostructure exhibit even and odd two different SOTs. The mechanism of this SOT has another origin rather than spin Hall and Rashba-Edelstein effects. Recently, several approaches to induce SOT have been proposed including orbital Hall effect [189–191], thermal gradients [191] and magnons [191,192]. The detail of discussion for these effects can be found in recent road map [191]. These effects are still under investigated. Nevertheless, advantages of 2D materials, which expose their interfaces, can provide opportunities to understand the SOT at the interfaces through the use of surface characterization techniques. Many more interesting and unknown mechanism behind SOT magnetization switching can be discovered in the near future by using 2D quantum materials.

**Acknowledgments**

This work was supported by the Institute for Basic Science of Korea (Grant No. IBS-R011-D1).

**Author contributions**

All authors contributed to the drafting of this manuscript. R. C. Sahoo and Dinh Loc Duong contributed equally in this work.

**Correspondence** should be addressed to PNH (Email: pham.n.ab@m.titech.ac.jp) and YHL (Email: leeyoung@skku.edu).

**Competing interests statement**

The authors declare no competing interests.

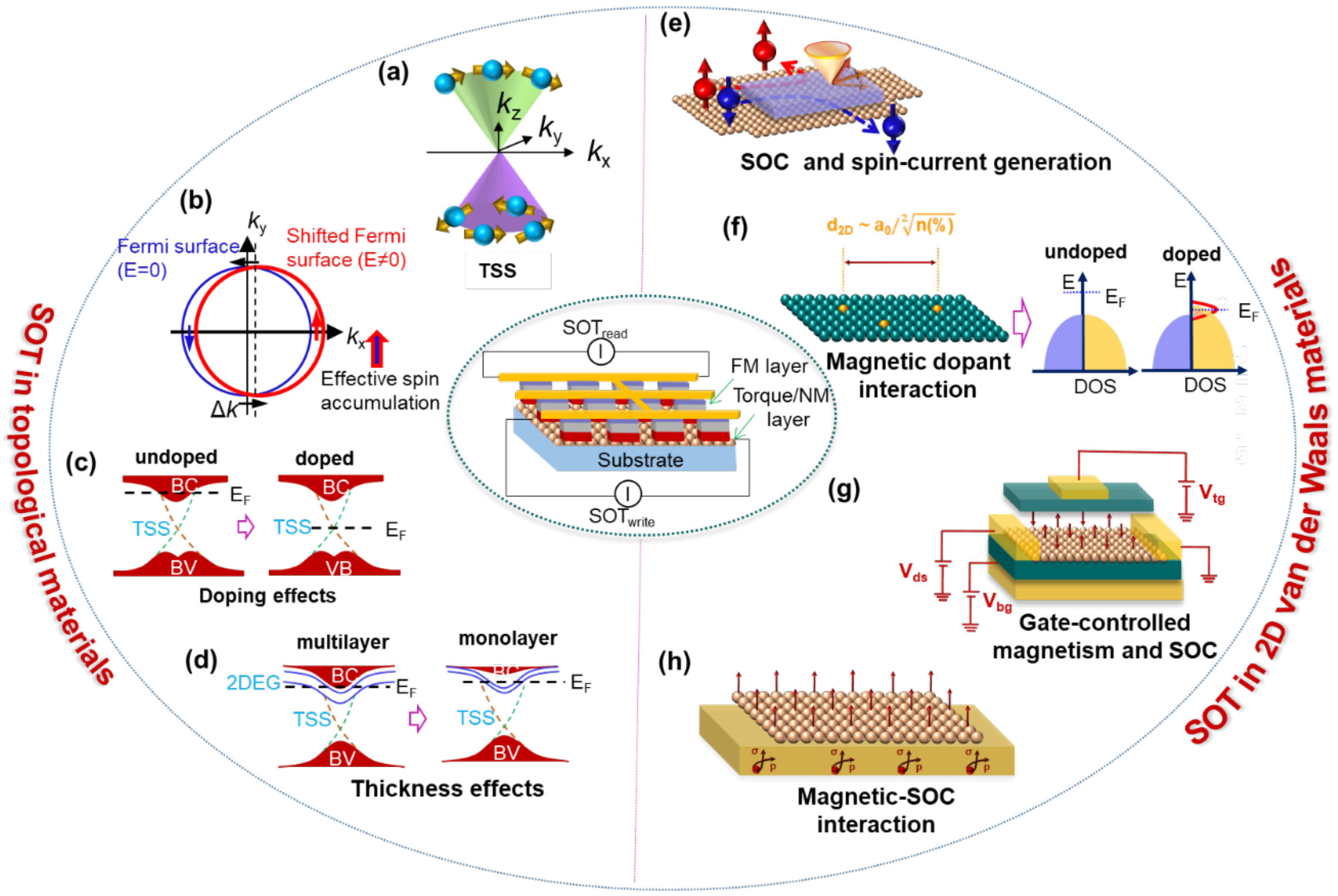

**Fig. 1** Opportunities in topological insulators and two-dimensional layered materials. **a** Dirac cone of the TI surface state with helical spin texture. The spin and momentum are locked at right angles to each other in the topologically protected surface states. **b** Bottom view of the Fermi surfaces on the $k_x - k_y$ plane. An applied charge current along $k_x$ generates an effective net momentum of the topological surface states (TSSs) in the same direction and an effective surface spin accumulation along the $k_y$ direction due to the spin-momentum locking of the charge carriers. The arrows indicate the spin magnetic moment directions of the carriers. **c** Schematic of the Fermi levels ($E_F$) modulation with doping effect in TIs. At the critical doping compositions, the $E_F$ are in the bulk bandgap of TIs and approach toward the Dirac point. **d** Schematic of layer dependent band structure in TIs. **e** Spin-current generated by a charge-current in the proximity of strong spin-orbit coupling 2D vdWs materials. **f-h** Owing to the nature of the atomically thick layer, the properties of 2D layered vdWs materials can be easily tuned by various means such as external electric fields, strain, different magnetic dopant, and proximity with other materials. 2D magnetic materials also reveal the layer-dependent magnetic exchange coupling between the layers.

The magnetic interaction between dopants in 2D materials may also differs from their 3D configurations.

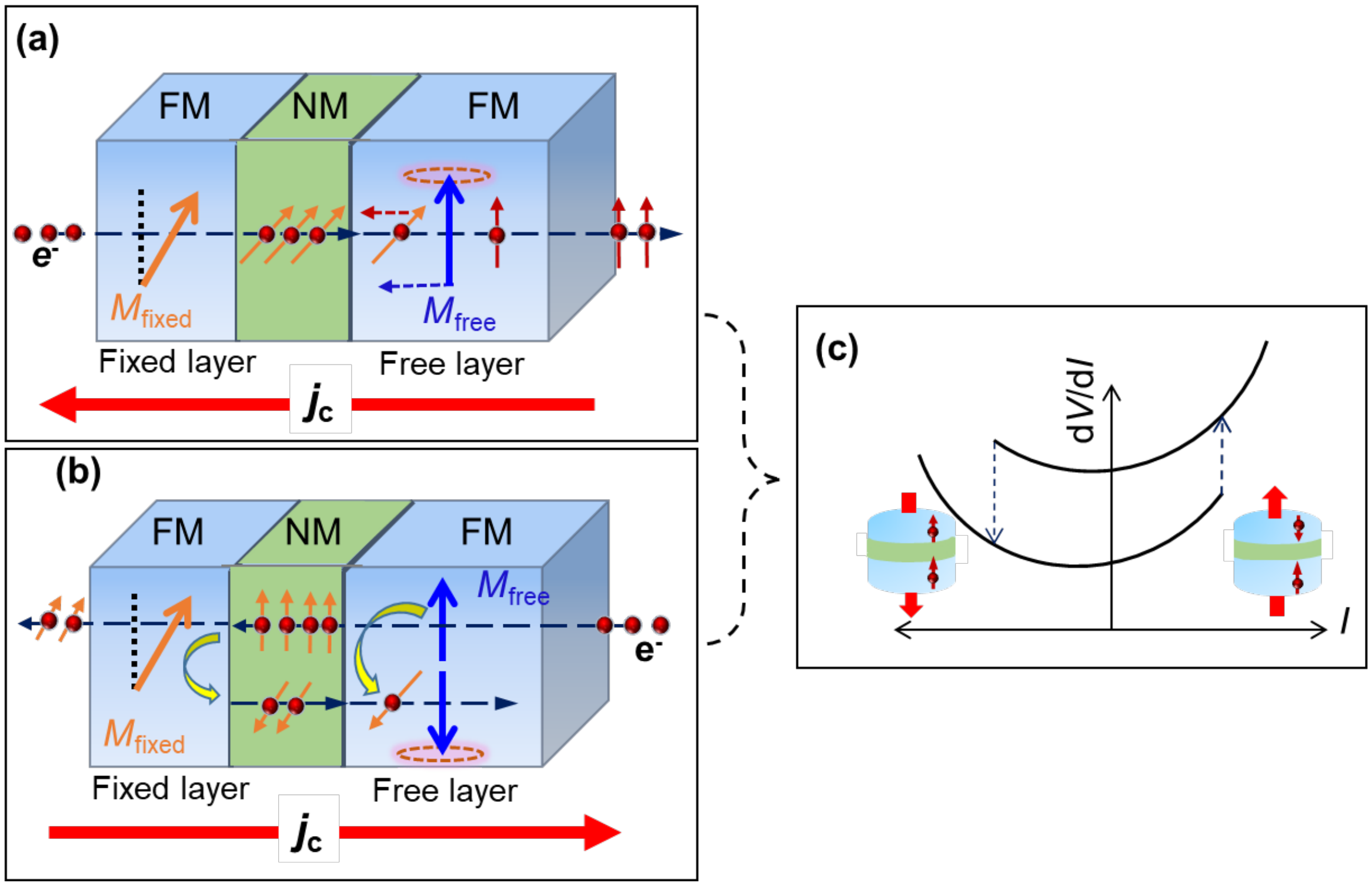

**Fig. 2** Concept of spin-transfer torque mechanism. **a** Schematic of a conventional spin-transfer torque structure consisting of fixed and free ferromagnetic layers ($FM_{fixed}$ and $FM_{free}$) sandwiching a nonmagnetic (NM) tunnel layer. $\boldsymbol{M}_{fixed}$ and $\boldsymbol{M}_{free}$ are the magnetizations of $FM_{fixed}$ and $FM_{free}$, respectively. The unpolarized conduction electrons are polarized in the direction of $\boldsymbol{M}_{fixed}$, and their transverse polarization components are absorbed by $\boldsymbol{M}_{free}$ when charge current ($\boldsymbol{j}_c$) passes through the $FM_{free}$/NM/$FM_{fixed}$ heterostructure. As a result, $\boldsymbol{M}_{free}$ becomes parallel to the $\boldsymbol{M}_{fixed}$ owing to the spin-torque. **b** A reverse current flow ($FM_{fixed} \rightarrow FM_{free}$) switches $\boldsymbol{M}_{free}$ to become antiparallel to $\boldsymbol{M}_{fixed}$. **c** Output *I*-*V* characteristics of a typical nanopillar STT device. Electron flow (positive bias) above the threshold current from the bottom $FM_{fixed}$ to the top $FM_{free}$ layer or vice versa can switch the magnetization direction and show different resistance states and hysteresis behaviours.

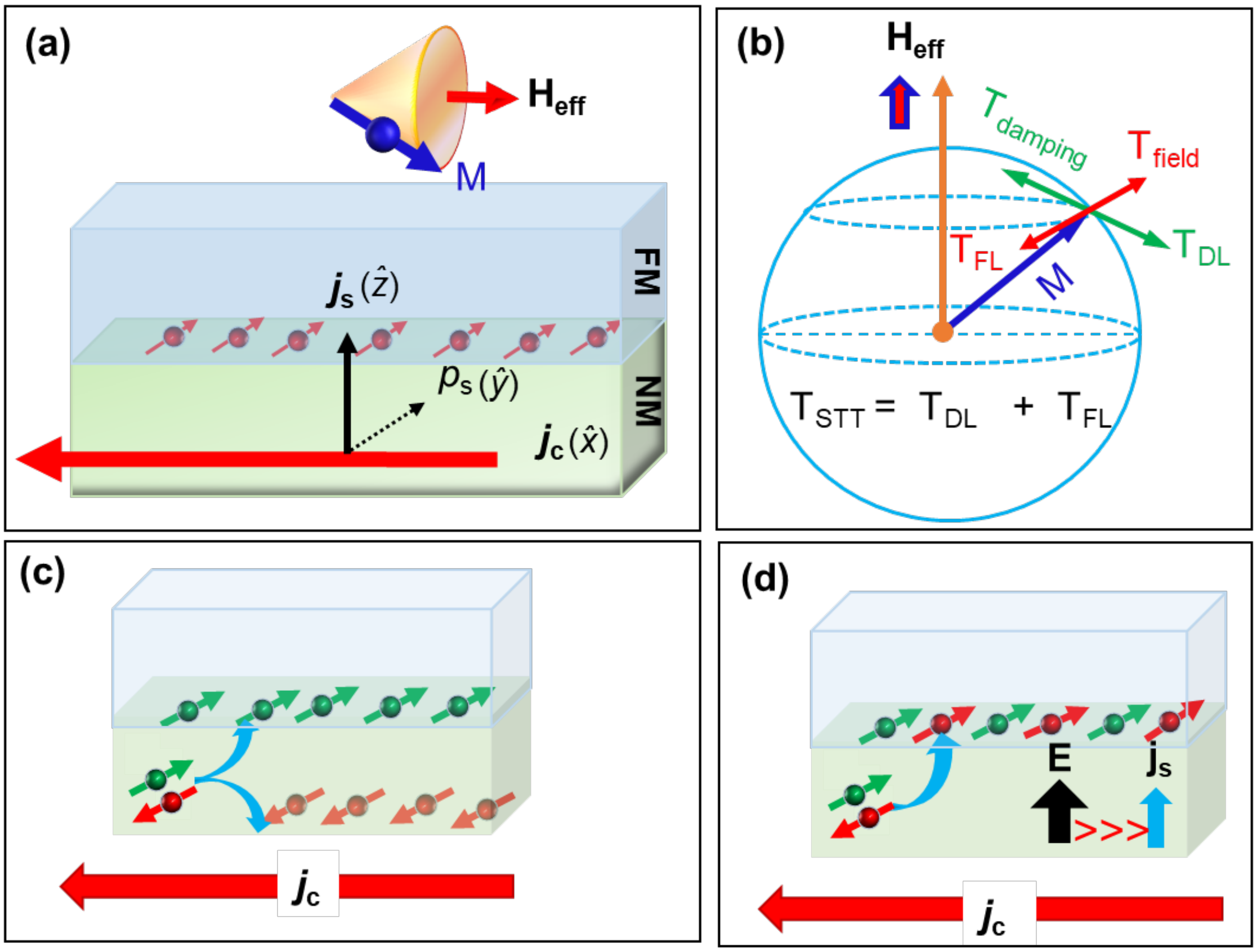

**Fig. 3** Concept and mechanism of spin-orbit torque. **a** Spin-orbit torque (SOT) device composed of an FM layer and an NM layer. A longitudinal $\boldsymbol{j}_c$ passing through the NM layer induces a transverse spin current ($\boldsymbol{j}_s$) due to the strong spin-orbit coupling of the NM layer. $\boldsymbol{j}_s$ flows into the FM layer and exerts a torque on the local magnetization of the FM layer that then precesses about the effective field ($\boldsymbol{H}_{eff}$). **b** The enlarged view of the total spin-torque acting on the local magnetization of FM. The total torque is the sum of the damping-like ($T_{DL}$, parallel or antiparallel to the damping) and field-like ($T_{FL}$, orthogonal to the plane shared by both FM magnetization and $H_{eff}$) torques. Two key mechanisms (**c** spin Hall effect and **d** Rashba-Edelstein effect) that induce SOT at the NM/FM interface.

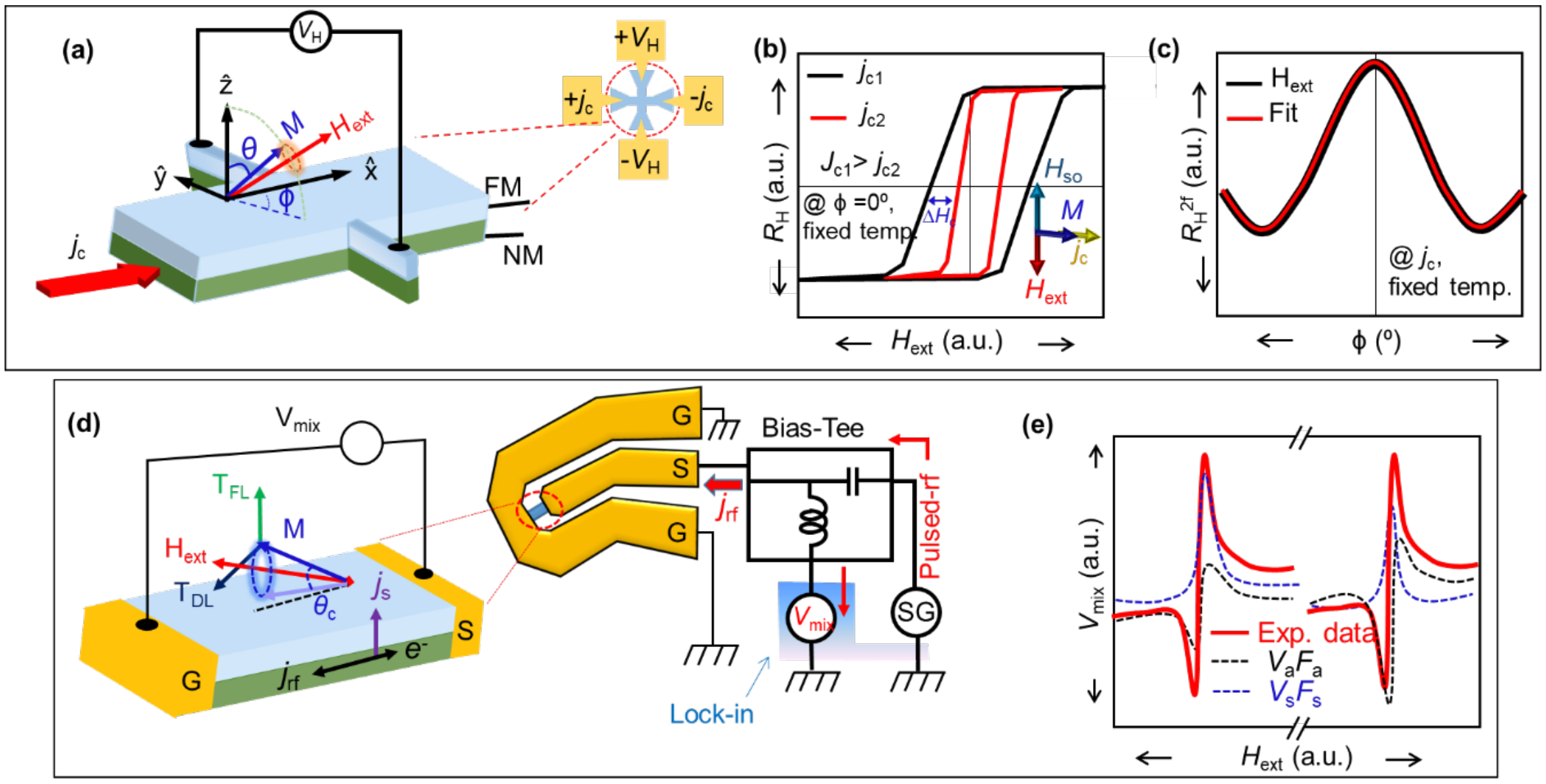

**Fig. 4** Electrical measurement method for probing SOT. **a** Hall-bar measurement geometry for SOT. Schematics of the experimental nanodevice structure with conventional contact pads. **b** External magnetic field ($H_{ext}$) dependence of DC Hall resistance ($R_H$). **c** Second-harmonic Hall resistance ($R_H^{2f}$) as a function of applied field angle (ϕ°) with different $H_{ext}$. **d** Schematics of the experimental geometries for spin-torque ferromagnetic resonance (ST-FMR). The bilayer sample is integrated into a terminated coplanar waveguide along with Bias-Tee circuit. **e** Output ($V_{mix}$) of the ST-FMR measurement as a function of $H_{ext}$. The dotted lines are fitted to Lorentzian function, showing symmetric and antisymmetric resonance components.

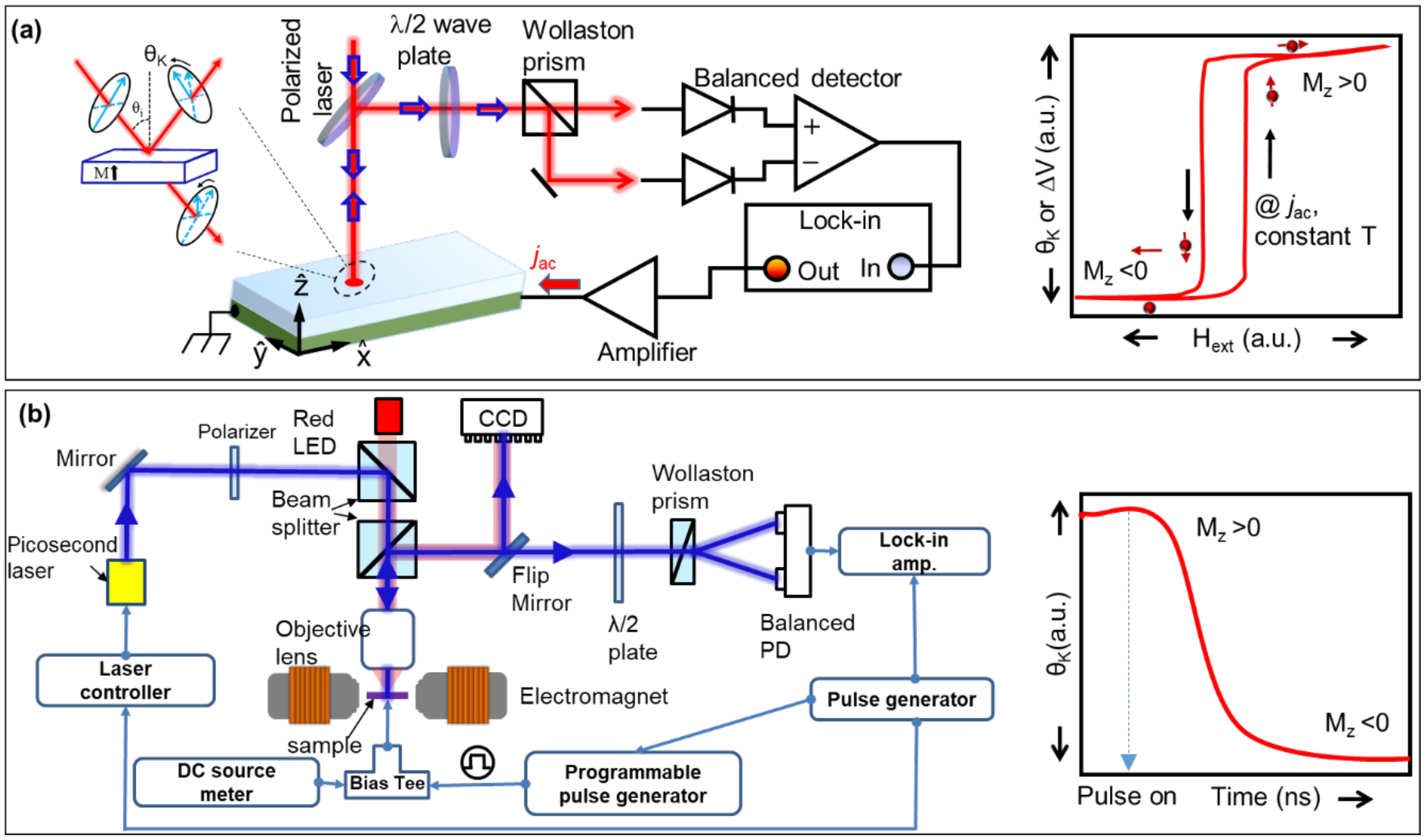

**Fig. 5** Optical measurement method for probing SOT. **a** Schematic illustration of the experimental setup of MOKE-based magnetometer with in-plane AC current $j_{ac}$ ($=j_{c0}\times\sin(ft)$) at frequency $f$ and scattering geometry of the probe laser beam. The change in the polarization of the reflected laser beam is analyzed using a λ/2-wave plate at 45°, Wollaston prism (beam splitter) and balanced detector. Lock-in amplifier is used to measure the Kerr angle $\theta_K$ or voltage change $\Delta V$ by $j_{ac}$ with the presence of magnetic field. Output of the lock-in amplifier from MOKE measurement as a function of $H_{ext}$ is depicted in the right panel. **b** The time-resolved MOKE setup for SOT-driven magnetization switching and domain-wall-motion measurement. Corresponding output is shown in the right panel.

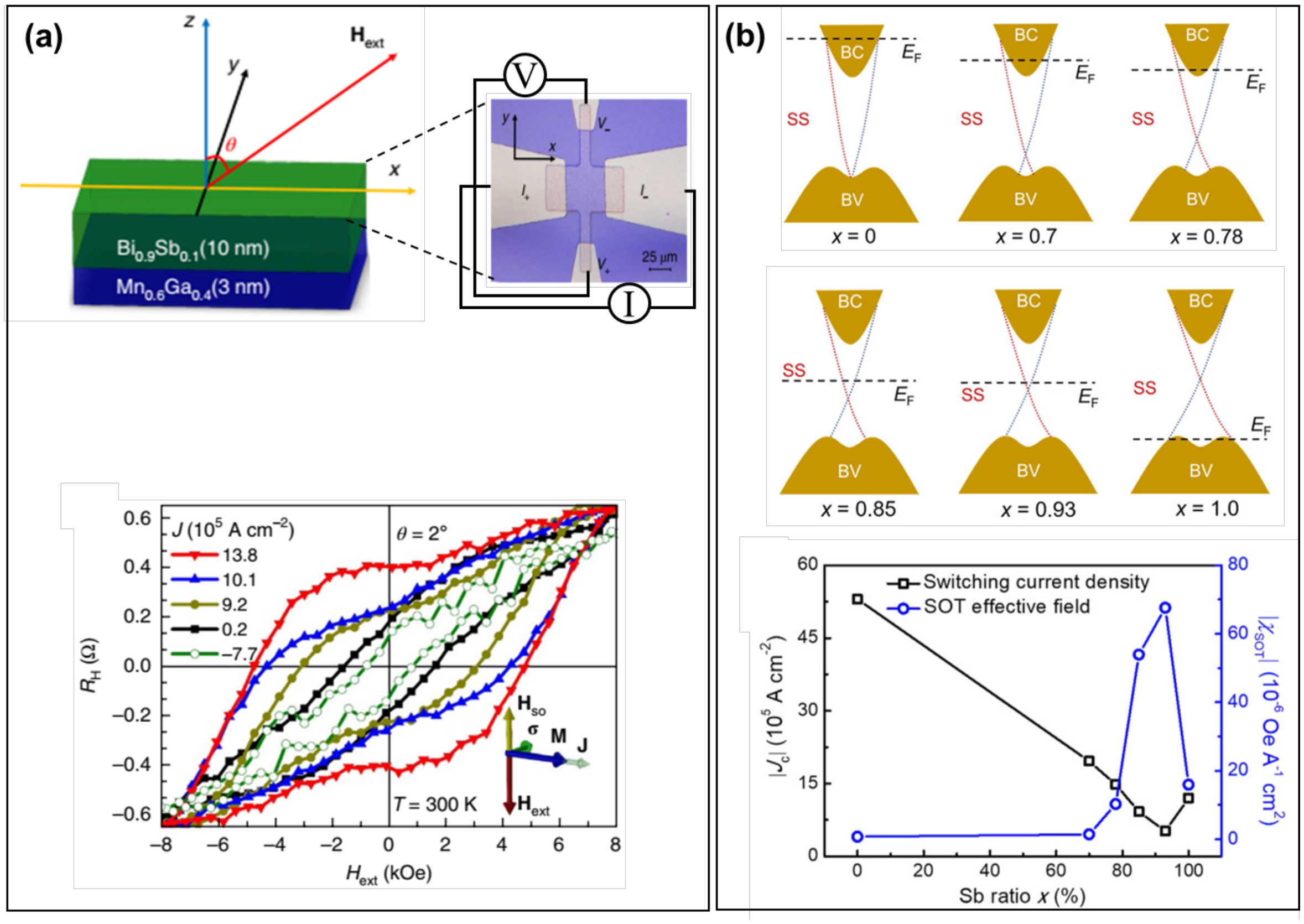

**Fig. 6** Topological materials for SOT. **a** Left panel: a 3D schematic of a SOT device comprising $Bi_{0.9}Sb_{0.1}$ (10 nm) and $Mn_{0.6}Ga_{0.4}$ (3 nm) layers. $H_{ext}$ is the externally applied magnetic field on the $zx$-plane at an angle $\theta$ to the $z$-axis. The yellow arrow along the $x$-axis represents the applied electric current direction. Right panel: a top view of the Hall bar device structure of the same heterostructure used for the corresponding SOT measurements. Lower panel: room temperature DC Hall resistance ($R_H$) of the $Bi_{0.9}Sb_{0.1}$ (10 nm)/$Mn_{0.45}Ga_{0.55}$ (3 nm) Hall bar device with a small tilted magnetization ($\boldsymbol{M}$) as a function of $H_{ext}$, measured at $\theta = 2°$ and different $j_c$ ($13.8 \times 10^5$ to $-7.7 \times 10^5$ A/cm$^2$). (Inset) a macrospin model where the spin-orbit field ($H_{so}$) is perpendicular to $\boldsymbol{M}$ at $H_{ext}$=-$H_c$ (coercive magnetic field at $j_c$). Reprinted figures are taken from Ref.[82]. **b** Upper panel: Schematic of the Fermi levels ($E_F$) at different Sb compositions in $(Bi_{1-x}Sb_x)_2Te_3$ (x = 0, 0.7, 0.78, 0.85, 0.93, 1.0) TIs. When the Sb compositions are about 85% and 93%, the $E_F$ is in the bulk bandgap of $(Bi_{1-x}Sb_x)_2Te_3$ and approach the Dirac point. Lower panel: | $j_c$ | (switching current) and | $\chi_{SOT}$ | (SOT-driven effective field) as a function of Sb ratio. This

implies that both insulating bulk and conducting surface states are responsible for large $\chi_{SOT}$ and small $j_c$ in $(Bi_{1-x}Sb_x)_2Te_3$. Reprinted figures are taken from Ref.[32]

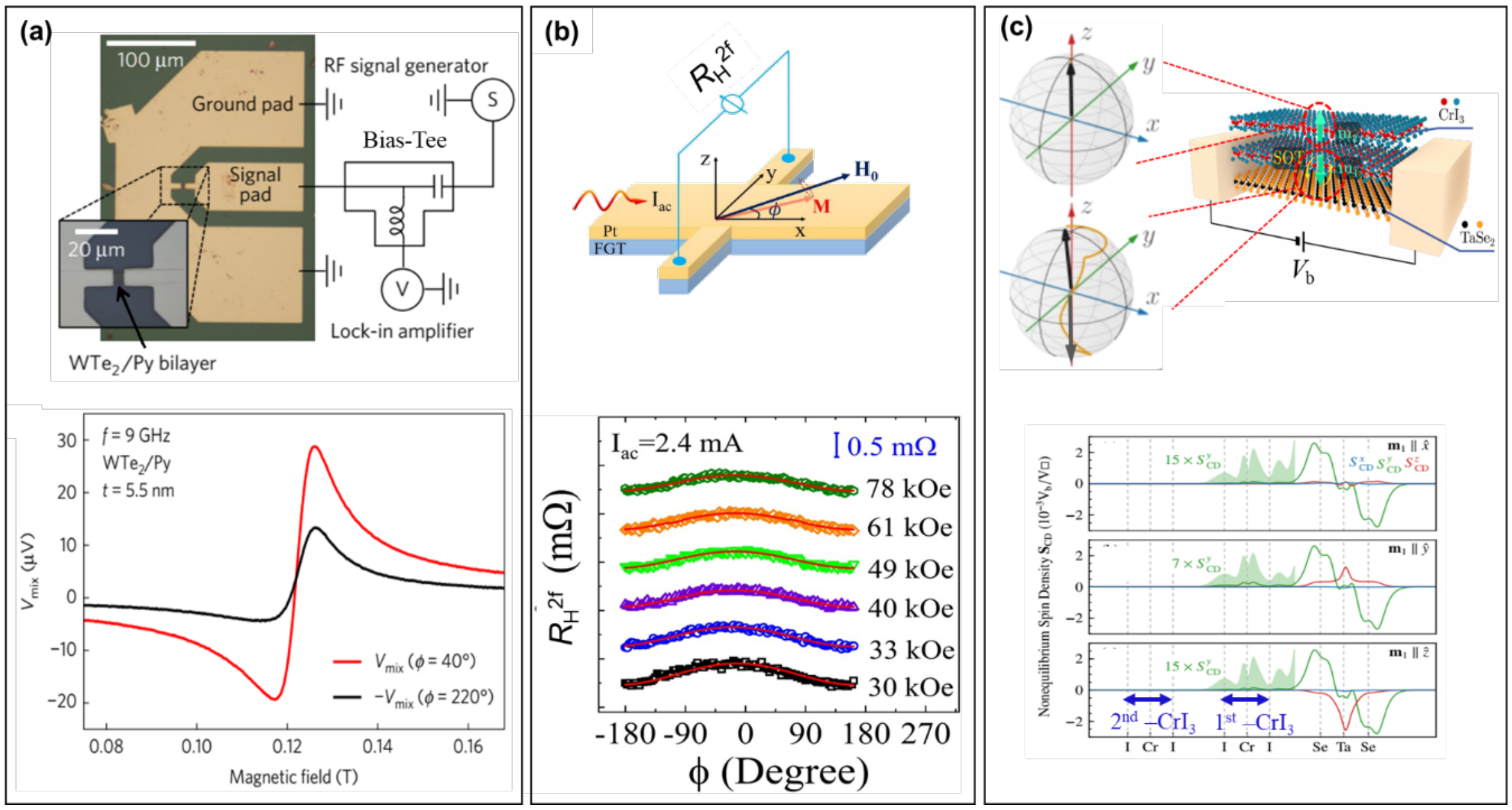

**Fig. 7** Two-dimensional materials for SOT. **a** Upper panel**:** Micrograph of the $WTe_2$ (5.5 nm)/Py (6 nm) heterostructure device with contact pads and ST-FMR measurement setup. Lower panel: Output resonances of ST-FMR measurements for the SOT heterostructure at RT and a frequency of 9 GHz with two different magnetization directions rotated by 180° (e.g. $\phi$ = 40° and 220°) with respect to the applied current direction. Reprinted figures are taken from Ref.[38]. **b** Upper panel: Schematic device structure of a $Fe_3GeTe_2$ (15–23 nm)/Pt (5 nm) bilayer and the corresponding coordinate system used for second-harmonic Hall signal measurements, where $M$ is the magnetization of $Fe_3GeTe_2$, $I_{ac}$ is the in-plane injected AC current, and $H_0$ is the in-plane external magnetic field or $H_{ext}$. Lower panel: the second-harmonic Hall resistance $R_H^{2f}$ of a $Fe_3GeTe_2$ (23 nm)/Pt (5 nm) bilayer as a function of $\phi$ (azimuthal angle between the applied AC current ($I_{ac}$=2.4 mA) and $H_{ext}$ direction) recorded at different fixed $H_{ext}$. Reprinted figures are taken from Ref. [11]. **c** Upper panel: Bilayer-$CrI_3$/monolayer-$TaSe_2$ hybrid device for SOT switching. In the presence of a small bias voltage ($V_b$), an unpolarized current through the monolayer-$TaSe_2$ generates a SOT strong enough to switch the magnetization ($\mathbf{m}_1$) of the first $CrI_3$ layer, whereas the magnetization ($\mathbf{m}_2$) of the second $CrI_3$ layer remains unchanged because of the zero-spin density in this layer, which is clearly observed in the current-induced non-equilibrium spin density ($S_{CD}$=$S_{CD}^{x,y,z}$) in the linear response region (lower panel). Reprinted figures are taken from

Ref. [118].

**Table 1**

Comparison of spin Hall angle, $\theta_{SH}$; conductivity, $\sigma$; spin Hall conductivity, $\sigma_{SH}$; power consumption, *P*; and operating temperature of various torque layers.

| | | Critical Current or $j_c$ (A $cm^{-2}$) | SOT efficiency or $\|\theta_{SH}\|$ | $\sigma$ ($\Omega^{-1}cm^{-1}$) | $\sigma_{SH}=(\hbar/2e)(\theta_{SH}\times\sigma)$ ($\hbar/2e\ \Omega^{-1}cm^{-1}$) | $P \propto 1/(\sigma\times\theta_{SH}^2)$ (a.u.) | Working temperature |
|---|---|---|---|---|---|---|---|
| Light metal | Cu(O) [121] | - | 0.08 | $1.2\times10^4$ | $0.96\times10^3$ | $0.13\times10^{-1}$ | RT |
| Heavy metal | β-Ta [107] | $5.5\times10^6$ | 0.15 | $5.3\times10^3$ | $0.79\times10^3$ | $0.83\times10^{-2}$ | RT |
| | β-W [110] | $3.2\times10^9$ | 0.33 | $3.8\times10^3$ | $1.25\times10^3$ | $0.24\times10^{-2}$ | RT |
| | W(O) [110] | $4.57\times10^6$ | 0.49 | $6.0\times10^3$ | $2.94\times10^3$ | $0.69\times10^{-3}$ | RT |
| | Pt [23,84] | $3.35\times10^7$ | 0.013 - 0.16 | $5.0\times10^4$ | $2.50\times10^3$ | $0.80\times10^{-2}$ | RT |
| Alloy | $Cu_{72}Pt_{28}$ [122] | - | 0.054 | $1.6\times10^4$ | $0.86\times10^3$ | $0.21\times10^{-1}$ | RT |
| | $Cu_{99.5}Bi_{0.5}$ [123] | - | 0.24 | $1.0\times10^5$ | $0.24\times10^4$ | $0.17\times10^{-3}$ | 10 K |
| | $Cu_{97}Ir_3$ [124] | - | 0.027 | $7.1\times10^4$ | $1.92\times10^3$ | $0.19\times10^{-1}$ | 10 K |
| Rashba interface | STO/LAO [36] | $10^5$ | 6.3 | $1.1\times10^2$ | $0.68\times10^3$ | $0.23\times10^{-3}$ | RT |
| | Bi/Ag [106] | - | 0.18 | $1.7\times10^5$ | $3.06\times10^4$ | $0.18\times10^{-3}$ | RT |
| Antiferromagnet | PtMn [105] | $1.0\times10^9$ | 0.10 | $4.4\times10^3$ | $0.4\times10^3$ | $0.23\times10^{-1}$ | RT |
| | IrMn [106] | - | 0.60 | $1.2\times10^4$ | $7.2\times10^3$ | $0.23\times10^{-3}$ | RT |
| Topological insulator | $Bi_2Se_3$ [30] | $2.8\times10^6$ | 2-3.5 | $(5.5\text{-}5.7)\times10^2$ | $(1.1\text{-}2)\times10^3$ | $(0.45\text{-}0.14)\times10^{-3}$ | RT |
| | $(Bi, Sb)_2Te_3$ [108] | $2.5\times10^6$ | 0.40 | $2.5\times10^2$ | $0.1\times10^3$ | $0.25\times10^{-1}$ | 2 K |
| | $Bi_xSe_{1-x}$ [142] | $4.3\times10^5$ | 18.8 | $7.8\times10^1$ | $1.47\times10^3$ | $0.36\times10^{-4}$ | RT |
| | $(Bi_{0.5}Sb_{0.5})_2Te_3$ [31] | - | 25 | $1.7\times10^3$ | $4.25\times10^4$ | $0.94\times10^{-6}$ | <200 K |
| | $(Bi_{0.5}Sb_{0.5})_2Te_3$ [128] | $8.9\times10^4$ | 140-410 | $2.2\times10^3$ | $(3.08\text{-}9.02)\times10^5$ | $(2.3\text{-}0.27)\times10^{-8}$ | 1.9 K |
| | $Bi_{0.9}Sb_{0.1}$ [82] | $1.5\times10^6$ | 52 | $2.5\times10^3$ | $1.30\times10^5$ | $0.48\times10^{-6}$ | RT |
| | $Bi_2Te_3$ [32,48] | $2.43\times10^6$ | 1.76 | $8.35\times10^2$ | $1.47\times10^3$ | $3.86\times10^{-4}$ | RT |
| | $Bi_xSe_{1-x}$/Ta [142] | $2.0\times10^7$ | 6 | - | - | - | RT |
| | Cr-$Bi_xSb_{2-x}Te_3$ [15] | $2.57\times10^5$ | 0.3-90 | - | - | - | RT-2.5 K |
| 2D material | $MoS_2$ [39] | - | 0.14 | $2.1\times10^2$ | $2.88\times10^1$ | $0.25\times10^{-2}$ | RT |
| | $WTe_2$ [38,102] | $2.96\times10^5$ | 0.013 | $6.1\times10^3$ | $8.0\times10^1$ | 0.97 | RT |
| | $WSe_2$ [39] | - | – | $6.3\times10^3$ | $5.52\times10^1$ | – | RT |
| | β-$MoTe_2$ [158] | - | 0.032 | $1.8\times10^3$ | $5.8\times10^1$ | 0.05 | RT |
| | $NbSe_2$ [159] | - | 0.005-0.013 | $(6\text{-}6.15)\times10^3$ | $(3\text{-}8)\times10^1$ | 6.6-0.96 | RT |
| | $PtTe_2$ [39] | $3.1\times10^5$ | 0.05-0.15 | $(0.3\text{-}3)\times10^4$ | $(0.2\text{-}1.6)\times10^3$ | $0.13\text{-}0.14\times10^{-2}$ | RT |
| | $TaTe_2$ [193] | - | - | $1.4\times10^4$ | $(10\text{-}20)\times10^1$ | - | RT |
| | $TaS_2$ [194] | $5.1\times10^5$ | 0.25 | $5.96\times10^4$ | $14.9\times10^3$ | $2.68\times10^{-4}$ | RT |